\documentclass[prb,amsfonts,amssymb,superscriptaddress]{revtex4}
\newcommand{\be}{\begin{equation}}
\newcommand{\ee}{\end{equation}}
\newcommand{\bea}{\begin{eqnarray}}
\newcommand{\eea}{\end{eqnarray}}
\newcommand{\tr}{\text{tr}}
\usepackage{graphicx}
\begin{document}

\title{Spin noise of itinerant fermions}
\author{\v Simon Kos}
\email{simonkos@kfy.zcu.cz}
\affiliation{Department of Physics, University of West Bohemia, Univerzitn\'\i \ 22,
306 14 Plze\v n, Czech Republic}
\affiliation{Cavendish Laboratory, Cambridge University, Madingley Road, 
Cambridge, CB3 0HE, United Kingdom}
\affiliation{Theoretical Division, Los Alamos National Laboratory,
Los Alamos, New Mexico 87545, USA}
\author{
Alexander V. Balatsky}
\affiliation{Theoretical Division, Los Alamos National Laboratory,
Los Alamos, New Mexico 87545, USA}
\affiliation{Center for Integrated Nanotechnologies, Los Alamos, NM 87545, USA}
\author{
Peter B. Littlewood}
\affiliation{Cavendish Laboratory, Cambridge University, Madingley Road, 
Cambridge, CB3 0HE, United Kingdom}
\author{
Darryl L. Smith}
\affiliation{Theoretical Division, Los Alamos National Laboratory,
Los Alamos, New Mexico 87545, USA}

\begin{abstract}
We develop a theory of spin noise spectroscopy of itinerant, noninteracting,
spin-carrying fermions in different regimes of temperature and disorder.
We use kinetic equations for the density matrix in spin variables.
We find a general result with a clear physical interpretation, and discuss
its dependence on temperature, the size of the system, and applied magnetic field.
We consider two classes of experimental probes: 1. electron-spin-resonance
(ESR)-type measurements, in which the probe response to a uniform magnetization
increases linearly with the volume sampled,
and
2. optical Kerr/Faraday rotation-type measurements, in which the probe
response to a uniform magnetization increases linearly with the length of
the light propagation in the sample, but is independent of the cross
section of the light beam.
Our theory provides a framework for interpreting recent experiments on
atomic gases and conduction electrons in semiconductors and provides a baseline
for identifying  the effects of interactions on
spin noise spectroscopy.
\end{abstract}
\maketitle

\section{Introduction}

Currently, there is much interest in studying the physics of nano-scale
structures.
In measurements of a response function by
pump-probe experiments on systems of decreasing size,
the signal decreases more rapidly than
the noise, and thus the signal-to-noise ratio decreases with decreasing system size. The
fluctuation-dissipation theorem, which relates a response function to a correlation function obtained
from noise measurements, enables us to change this problem into a
useful tool. An additional advantage of noise measurements is that they often
disturb the system less than experiments that measure the response of the
system to an external perturbation.

There have been a number of experiments studying spin properties of systems
using spin noise.
Spin noise has been measured in systems of spins whose position is
{\it fixed in space}: atomic nuclei \cite{Sleator}, spin glasses \cite{Weissman},
magnetization modes in magnetoresistive heads \cite{Smith}, and electrons and
holes in self-assembled quantum dots \cite{dot}.
There have also
been recent measurements of spin noise of {\it itinerant} spins in hot atomic gases
\cite{Aleksandrov,Mitsui,noise nature}, cold atomic gases \cite{Sorensen,Kuzmich},
in $n$-doped bulk GaAs \cite{Oestreich,nGaAsbulk},
and in $n$-doped GaAs quantum wells \cite{GaAsQW}.  Localized spin noise measurements on nanostructured systems using STM techniques have been discussed \cite{STM1,STM2}.  The experimental setup of spin-noise
spectroscopy in semiconductors has been optimized in Ref. \cite{review} and
has been used to measure spatially resolved doping concentration
in GaAs \cite{doping}.
Motivated by these experiments with itinerant spins, we have developed
a theory of spin noise of itinerant
fermions in different regimes
of temperature (degenerate/classical statistics)
and disorder (ballistic/diffusive motion).
We consider the case of noninteracting particles as a benchmark
for comparison to experiments, so that we can then identify the effects of
interactions on spin noise. We find a general result that holds in the
different regimes, which has a clear physical interpretation, and we show
how it follows from kinetic equations for density matrix in spin
variables.
We consider two classes of experimental probes: 1. electron-spin-resonance
(ESR)-type measurements, in which the probe response to a uniform magnetization
increases linearly with the volume sampled,
and
2. optical Kerr/Faraday rotation-type measurements, in which the probe
response to a uniform magnetization increases linearly with the length of
the light propagation in the sample, but is independent of the cross
section of the light beam.

The outline of the paper is as follows: In Sec. \ref{General result}, we
present the general result and its interpretation. In Sec. \ref{Implications},
we show how the noise spectrum behaves as a function of temperature, system size, and
magnetic field. In Sec. \ref{Calculation}, we present the derivation of
the result from kinetic equations with details of calculations in some
limiting cases presented in the Appendix B. In Sec. \ref{Conclusions}, we
present our Conclusions. The spin noise power spectrum of spin $J$, which
motivates our general result, is derived in the Appendix A.

\section{General result}
\label{General result}
\begin{figure}
\includegraphics[width=0.35\textwidth,clip=]{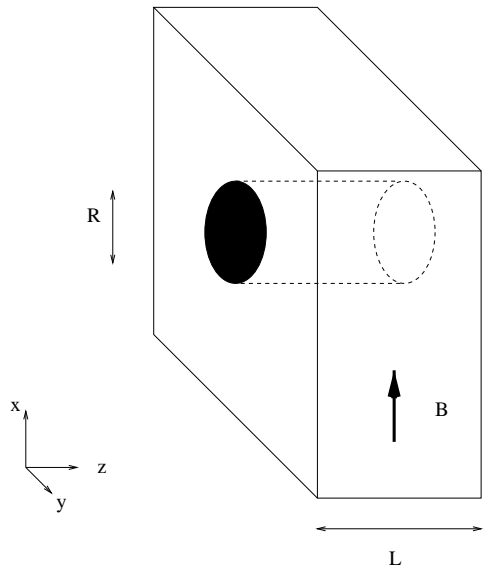}
\caption{Schematic view of the experimental setup. The experiment measures
the noise of spin magnetization in the $z$ direction in the presence of an
applied constant magnetic field in the $x$ direction that splits the two spin
energy levels by the Larmor frequency $\omega _L$. The system has thickness
$L$ in the $z$ direction, and is extended in the $x-y$ direction.
Noise in the
part of the system with transverse size $R$ and
cross section $A\sim R^2$ is probed.
\label{setup}}
\end{figure}
In Fig. \ref{setup} we show the generic setup of the experiments considered.
We study the
noise of spin magnetization in the $z$ direction in the presence of an
applied constant magnetic field in the $x$ direction that splits the two spin
energy levels by the Larmor frequency $\omega _L$. The purpose of the
constant magnetic field is to shift the
noise spectrum away from zero frequency. Noise measurements close to
zero frequency are difficult because of the presence of ubiquitous $1/f$ noise.
The magnetic field
is chosen so that $\omega _L$ is larger than the linewidth of
the spin noise spectrum.
For charged fermions, we neglect coupling of magnetic field to the orbital
motion, that is, we consider the case when the cyclotron orbit is longer
than the dimensions of the probed region.
The system has thickness
$L$ in the $z$ direction,
which in the optical experiments is the direction of light propagation.
The system is extended in the $x-y$ direction.
Noise in the
part of the system with transverse size $R$ and
cross section $A\sim R^2$ is probed.
For a uniformly spin-polarized sample, the magnitude of the
probe response in bulk measurements
such as ESR scales as the volume of the probed region, that is,
$
Signal \propto R^2 L.
$
By contrast, for a uniformly spin-polarized sample, the magnitude of the
probe response in optical Kerr/Faraday rotation measurements scales as
the thickness $L$, but is independent of the cross-sectional area $A$, that is,
$
Signal \propto L.
$

In general, the spin noise experiment will measure the noise power spectrum
of a quantity $Q$ proportional to the instantaneous electron spin polarization
\be
\label{geometric factor}
Q(t)=C M_z(t)(GF),
\ee
where $M_z(t)$ is the operator of the $z-$component of the instantaneous electron spin polarization in the probed volume at time $t$, related to the spin-density operator
$s_z({\bf r},t)$ by
\be
M_z(t) = \int\limits_{A\times L}d^3r s_z({\bf r},t),
\ee
$C$ is a fixed coupling constant, and
$(GF)$ is a geometric factor. For bulk measurement such as ESR, the geometric
factor is unity. For optical Kerr/Faraday rotation measurements, the geometric
factor is $1/A$. Hence, in either case, in order to calculate the spin noise
power spectrum, we need the Fourier transform of the correlation function
\be
\label{def of noise correlator}
S_{zz} (t_2-t_1)={1\over 2}\langle \{ M_z(t_2),M_z(t_1) \} \rangle,
\ee
where
Here, $\{,\}$ denotes the anticommutator, and $\langle \rangle$ is the
equilibrium ensemble average.

In equilibrium, the fluctuation-dissipation theorem relates the noise power to the imaginary
part of the corresponding susceptibility
\be
\label{FDthm}
S_{zz}(\omega ) = -\coth {\omega \over 2T}
\int\limits_{A\times L}d^3r _2 d^3 r _1
\chi _{zz}''({\bf r}_2-{\bf r}_1; \omega)
\simeq
-\coth {\omega \over 2T} A\times L
\chi _{zz}''(q=1/R; \omega)
,
\ee
where
\be
\chi _{zz} ({\bf r}_2-{\bf r}_1; t_2-t_1)
=-i\theta (t_2-t_1) \langle [s_z({\bf r}_2,t_2),s_z({\bf r}_1,t_1)]\rangle .
\ee
Here, $\chi _{zz}$ is the spin susceptibility.

\begin{figure}
\includegraphics[width=0.5\textwidth,clip=]{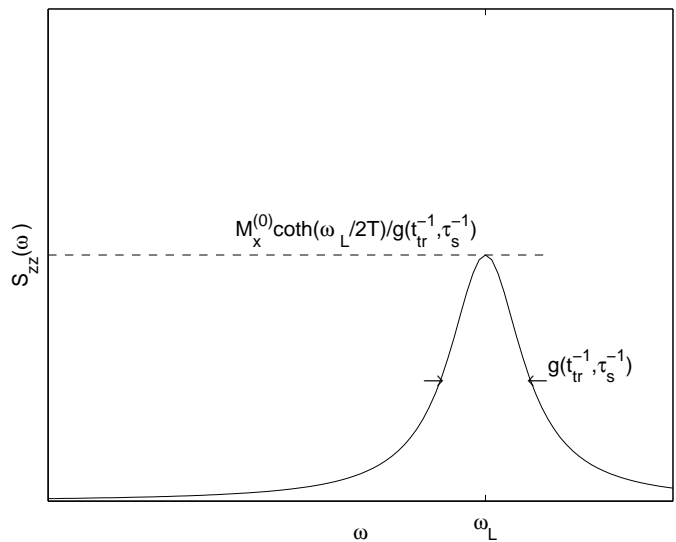}
\caption{General qualitative form of the noise power spectrum.
It is peaked at the
Larmor frequency $\omega _L$. Its width is approximately equal to the larger of the
inverse travel time $t_{tr}^{-1}$ and the inverse spin-flip time
$\tau _s^{-1}$.
The height is given by
the magnetization $M_x^{(0)}$ multiplied by the thermal factor
$\coth \omega _L/2T$ and divided by the width.
\label{figure with main result}}
\end{figure}
We consider different regimes of temperature and particle motion.
The calculations for the various cases are given in Sec. \ref{Calculation}.
The results of the calculation
in all these cases have the following
general form
\be
\label{general formula}
S_{zz}(\omega ) \simeq {\pi \over 2 } \coth {\omega _L \over 2T}
{M_x^{(0)} \over g (t_{tr}^{-1}, \tau _s^{-1})}
f\left( {\omega - \omega _L \over g (t_{tr}^{-1}, \tau _s^{-1})} \right).
\ee
Here, $M_x^{(0)}$ is the equilibrium magnetization of the probed region caused by the constant
magnetic field in the $x$ direction, $t_{tr}$ is the travel time it takes
the spin-carrying fermion to move across the probed region (distance $R$),
$\tau _s$ is the spin-flip time, $f$ is a function of
unit height and width, peaked at zero, and $g$ is a function whose value is
approximately equal to the greater of the two arguments.
The detailed form of the functions $f$ and $g$ depends on temperature and
disorder.
The form of the functions $f$ and $g$ for the various
cases will be discussed in section
\ref{Calculation}. Equation \ref{general formula} is the main result of this
paper. Its form agrees with the spin noise power spectrum of a single
spin $J$--see Appendix A.

In Eq. \ref{general formula},
for ballistic transport,
\be
t_{tr}=R/v,
\ee
where $v$ is the Fermi velocity $\sqrt{2E_F/m}$
in the degenerate regime $T<<E_F$ or the
thermal velocity $\sqrt{2T/m}$ for $T>>E_F$. For diffusive transport,
\be
t_{tr}=R^2/D,
\ee
where $D$ is the diffusion constant.

In Fig. \ref{figure with main result}, we show the schematic behavior
of the spin noise power spectrum $S_{zz}(\omega )$. It is peaked at the
Larmor frequency $\omega _L$. Its width is approximately equal to the larger of the
inverse travel time $t_{tr}^{-1}$ and the inverse spin-flip time
$\tau _s^{-1}$.
The height is given by
the magnetization $M_x^{(0)}$ multiplied by the thermal factor
$\coth \omega _L/2T$ and divided by the width.

The result (\ref{general formula}) has a simple physical interpretation. The
scale of the response in the $z$ direction is set by the initial polarization
in the $x$ direction. The width is given by the inverse of the time
to lose spin coherence, either due to a spin flip or by moving out of the probed
region. This inverse time divides the magnetization to give
the noise power the dimension of inverse frequency. The peak in the noise
power spectrum is
centered at the Larmor frequency $\omega _L$.

\section{Implications}
\label{Implications}

We discuss consequences of Eq.
(\ref{general formula}). Because $M_x^{(0)}$ grows linearly with the probed
volume, and the other quantities in Eq. (\ref{general formula}) are
independent of $L$, the height of the power
spectrum grows linearly with $L$ while its width is
independent of $L$.

\begin{figure}
\includegraphics[width=0.7\textwidth,clip=]{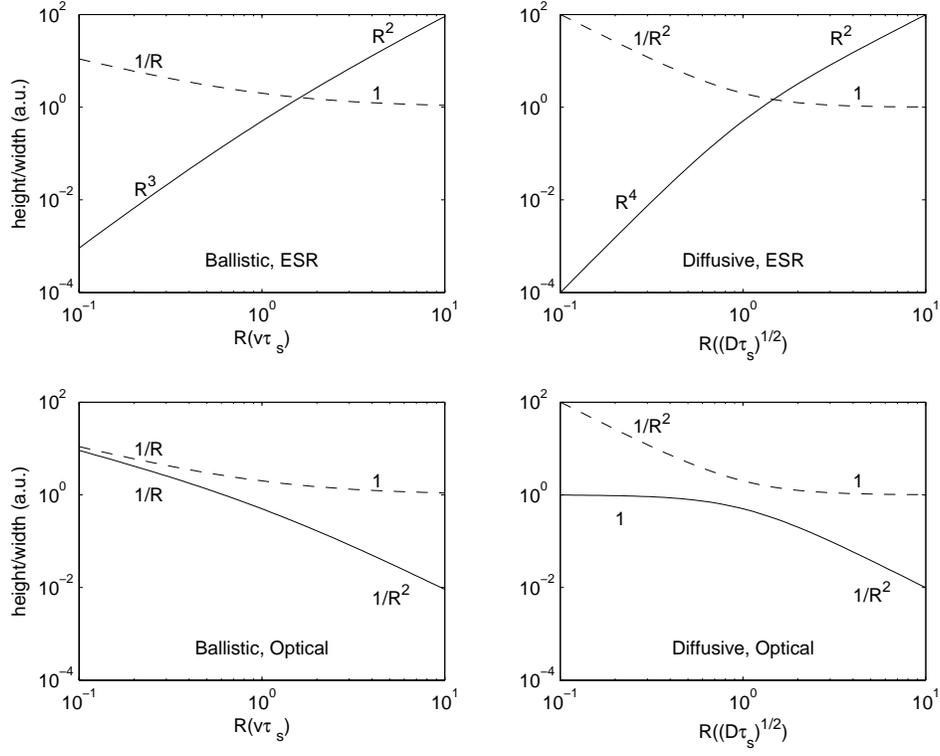}
\caption{Scaling of the height (solid line) and width (dashed line)
of the noise power spectrum as a function of $R$ in the ESR measurements (first row)
and optical measurements (second row). The first row are the log-log
plots of formulas (\ref{ESR ballistic h}), (\ref{ESR ballistic w}) and
(\ref{ESR diffusive h}), (\ref{ESR diffusive w}); the second row are the
log-log plots of the same formulas with the height divided by $R^4$. The heights and widths
are normalized to their value at $R=v\tau _s$ for the ballistic case (the
first column) and at $R=\sqrt{D\tau _s}$ for the diffusive case (the second
column). The scaling in the different regions is denoted at each curve.
The crossover is most pronounced for the optical measurement in the diffusive
regime.
\label{scaling with R fig}}
\end{figure}

The dependence of the noise power spectrum on $R$ is different for the
ballistic and diffusive motion. In the ballistic
case, the height of the noise power spectrum behaves as
\be
\label{ESR ballistic h}
S_{zz}(\omega _L)\propto {R^2 \over {v\over R} + \tau _s^{-1}},
\ee
and its width behaves as
\be
g (t_{tr}^{-1}, \tau _s^{-1}) \propto \label{ESR ballistic w}
{v\over R} + \tau _s^{-1}.
\ee
Thus, for $R<v\tau _s$, the height of the noise power spectrum scales as $R^3$,
and its width
as $1/R$, whereas for $R>v\tau _s$, the height of the noise power spectrum
scales as $R^2$ and its
width is independent of $R$. In the diffusive regime, the height
of the noise power spectrum
behaves as
\be
\label{ESR diffusive h}
S_{zz}(\omega _L)\propto {R^2 \over {D\over R^2}+ \tau _s^{-1}},
\ee
and its width behaves as
\be
\label{ESR diffusive w}
g (t_{tr}^{-1}, \tau _s^{-1}) \propto {D\over R^2}+ \tau _s^{-1}.
\ee
Thus, for $R<\sqrt{D\tau _s}$, the height of the noise power spectrum
scales as $R^4$ and its
width as $1/R^2$, whereas for $R>\sqrt{D\tau _s}$, the height
of the noise power spectrum scales as
$R^2$ and its width is independent of $R$. For both the ballistic and
diffusive transport in the long-$R$ limit, we recover the behavior of
static spins flipping on the time scale $\tau _s$.
For optical Kerr/Faraday rotation experiments, there is an additional factor
$1/A^2$ in the height of the noise power spectrum for the geometrical factor,
Eq. \ref{geometric factor}.  The scaling behavior is summarized in Fig. \ref{scaling with R fig}.
The scaling crossover is most pronounced in optical measurements
of the noise in the diffusive regime.

\begin{figure}
\includegraphics[width=0.7\textwidth,clip=]{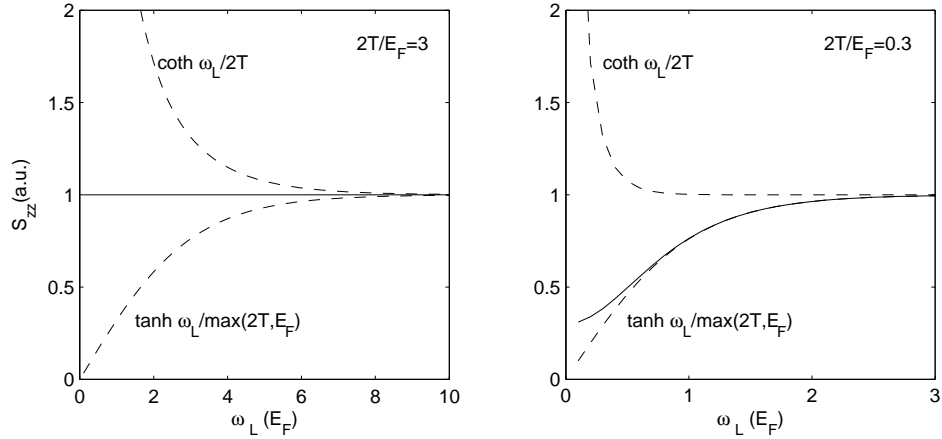}
\caption{Solid line: the magnetic-field dependence of the integrated noise
power spectrum $S_{zz}$ in the
classical ($2T/E_F=3$, left panel) and degenerate ($2T/E_F=0.3$, right panel)
regime. Dashed lines: the factors $\coth {\omega _L \over 2T}$
(upper dashed lines) and
$\tanh {\omega _L \over \max (2T, E_F)}$. The value of $S_{zz}$  is normalized to its
value in the high-field limit. In the classical regime, the field dependence of
the two factors cancels, and $S_{zz}$ is field independent. In the degenerate
regime, $S_{zz}$ grows linearly with field in the region $2T<\omega _L<E_F$.
\label{omega_L dependence fig}}
\end{figure}

We consider the dependence of the noise power spectrum
on the external magnetic field,
that is, on the Larmor frequency $\omega _L$. The width is independent
of $\omega _L$. Hence, instead of studying the width and the height
of the noise power spectrum separately, we study
the integrated spin noise power spectrum
\be
S_{zz} \equiv \int\limits_0^{\infty} d\omega S_{zz}(\omega),
\ee
which scales as the product of the  width and the height.
>From Eq. \ref{general formula}, we see that
\be
S_{zz} \propto  \coth {\omega _L \over 2T} M_x^{(0)},
\ee
that is, it is independent of the characteristic rate
$g (t_{tr}^{-1}, \tau _s^{-1})$.
In the classical regime, the
magnetization $M_x^{(0)}$ depends on $\omega _L$ as
\be
M_x^{(0)} \propto \tanh {\omega _L \over 2T}.
\ee
In the degenerate regime, $M_x^{(0)}$ also has a linear dependence on
$\omega _L$ for small $\omega _L$, cut off at the Fermi energy $E_F$. We,
therefore, approximate the field dependence of $M_x^{(0)}$ by
\be
M_x^{(0)} \propto \tanh {\omega _L \over \max (2T, E_F)},
\ee
giving
\be
\label{omega_L dependence formula}
S_{zz}\propto \coth {\omega _L \over 2T}
\tanh {\omega _L \over \max (2T, E_F)}.
\ee
In the classical regime, $2T/E_F>1$, the $\omega _L$-dependence of the two
factors in (\ref{omega_L dependence formula}) cancels, and $S_{zz}$
is independent of the magnetic field, in agreement with the measurements
of spin noise in Rb vapors
\cite{noise nature}.
In the degenerate regime, $2T/E_F<1$, there is an intermediate regime of
magnetic fields $2T<\omega _L<E_F$, where $S_{zz}$ grows
linearly with field, see Fig. \ref{omega_L dependence fig}. The spin noise
measurements in GaAs \cite{Oestreich,nGaAsbulk,GaAsQW} were done at temperature $T>>\omega _L$,
so $S_{zz}$ is field independent.

\section{Calculation}
\label{Calculation}
We now turn to a detailed justification of the above results.
We calculate the susceptibility $\chi _{zz}$ as a linear response of the
spin density $\langle s_z({\bf r},t) \rangle $ to an external potential
$\phi ({\bf r},t)$. The Hamiltonian describing the coupling is
\be
\label{H_phi}
H_{\phi }= \int d^3r \phi ({\bf r},t) s_z({\bf r},t)
.
\ee
In terms of the electron field operator in the Heisenberg representation,
$\psi ({\bf r},t)$, the spin density
operator $s_z({\bf r},t)$ is given as
\be
s_z({\bf r},t)=
\psi ^{\dag} ({\bf r},t) {\sigma _z \over 2} \psi ({\bf r},t),
\ee
where $\sigma _z$ is a Pauli matrix.
In order to calculate the linear response, we construct and solve the kinetic
equation for the density matrix in the Wigner representation
\be
\rho _{\alpha \beta}({\bf p},{\bf r},t)
=
\int d^3 \tilde{r} e^{-i{\bf p}\cdot {\tilde{\bf r}}}
\left \langle
\psi ^{\dag}_{\beta}
\left({\bf r}-{{\tilde{\bf r}}\over 2},t\right)
\psi _{\alpha}
\left({\bf r}+{{\tilde{\bf r}}\over 2},t\right)
\right \rangle .
\ee
In terms of this density matrix, the spin density is
\be
\langle s_z({\bf r},t) \rangle =
\int {d^3p \over (2\pi )^3}
\tr \left(\rho ({\bf p},{\bf r},t) {\sigma _z \over 2}\right).
\ee

The Hamiltonian consists of three terms: a non-interacting term
\be
\label{H_0}
H_0=\int d^3 r \left(
{1\over 2m}\nabla \psi ^{\dag}({\bf r},t) \nabla \psi ({\bf r},t)
-\omega _L \psi ^{\dag} ({\bf r},t) {\sigma _x \over 2} \psi ({\bf r},t)
\right),
\ee
a coupling term to the external potential (\ref{H_phi}), and a scattering
term, which determines whether the particle motion is ballistic or diffusive.
The first term on the right-hand side of Eq. \ref{H_0} is the kinetic
energy, and the second term describes the interaction with the applied magnetic
field. The kinetic equation for the density matrix has the following form
\be
\label{left-hand side formula}
\partial _t \rho _{\alpha \beta}({\bf p},{\bf r},t)
-i
\int d^3 \tilde{r} e^{-i{\bf p}\cdot {\tilde{\bf r}}}
\left \langle
\left[H_0+H_{\phi },
\psi ^{\dag}_{\beta}
\left({\bf r}-{{\tilde{\bf r}}\over 2},t\right)
\psi _{\alpha}
\left({\bf r}+{{\tilde{\bf r}}\over 2},t\right)
\right]
\right \rangle
= I_s
\ee
where $I_s$ describes the effect of scattering.
Substituting (\ref{H_phi}) and (\ref{H_0}) into (\ref{left-hand side formula}),
we find
\be
\left( \partial _t+ {{\bf p}\over m}\cdot \nabla \right)
\rho ({\bf p},{\bf r},t)
-i\omega _L \left[ {\sigma _x \over 2},\rho ({\bf p},{\bf r},t)\right]
-i
\left( \phi \left( {\bf r}-{i\over 2}\partial _{\bf p},t \right)
\rho ({\bf p},{\bf r},t) {\sigma _z \over 2}
-
\phi \left( {\bf r}+{i\over 2}\partial _{\bf p},t \right)
{\sigma _z \over 2} \rho ({\bf p},{\bf r},t) \right)
= I_s.
\ee
To obtain linear response, we write
\be
\rho ({\bf p},{\bf r},t) = \rho ^{(0)}(p) + \delta \rho ({\bf p},{\bf r},t).
\ee
Here, $\rho ^{(0)}(p)$ is the equilibrium density matrix corresponding to the
Hamiltonian $H_0$, that is,
\bea
\rho ^{(0)}(p)
& = &
{1\over 2}\left(n\left(\xi (p)-{\omega _L \over 2}\right)
+ n\left(\xi (p)+{\omega _L \over 2}\right)\right) \sigma _0
+
{1\over 2}\left(n\left(\xi (p)-{\omega _L \over 2}\right)
- n\left(\xi (p)+{\omega _L \over 2}\right)\right)  \sigma _x
\nonumber \\
& \equiv &
\rho ^{(0)}_0 (p) \sigma _0
+
\rho ^{(0)}_x (p) \sigma _x
.
\eea
Here
\be
n(\xi )= {1\over e^{{\xi - \mu \over T}}+1},
\ee
and
\be
\xi (p)={p^2 \over 2m}.
\ee
The particular deviations from equlibrium we consider have the form
\be
\label{deviations}
\delta \rho ({\bf p},{\bf r},t)
= \delta \rho _y ({\bf p},{\bf r},t) \sigma _y
+\delta \rho _z ({\bf p},{\bf r},t) \sigma _z
\equiv
\delta \mbox {\boldmath $\rho $} ({\bf p},{\bf r},t) \cdot
\mbox {\boldmath $\sigma $}.
\ee

The scattering contribution to the kinetic equation is different for the
ballistic and diffusive motion. We first consider the case of the ballistic
motion, in which
\be
I_s=-{1\over \tau _s} \delta \rho ({\bf p},{\bf r},t) .
\ee
It describes spin relaxation with relaxation time $\tau _s$.

To obtain the susceptibility at positive frequencies, we write
the kinetic equation for the circular component of the density matrix
\be
\delta \rho _+ ({\bf p},{\bf q},\omega ) =
\int d^3r dt e^{-i {\bf q}\cdot {\bf r} + i \omega t}
(\delta \rho _z -i \delta \rho _y) ({\bf p},{\bf r},t) .
\ee
The equation for the opposite circular component $\delta \rho _-$
gives the susceptibility at negative frequencies, which gives an equivalent
result.
The kinetic equation becomes
\be
-i\left( \omega -\omega _L-{{\bf p}\over m}\cdot
{\bf q} +{i\over \tau _s}\right)
\delta \rho _+  ({\bf p},{\bf q},\omega)
=i \phi ({\bf q},\omega)
{
\rho ^{(0)}_0 \left({\bf p}+{{\bf q}\over 2} \right)
-\rho ^{(0)}_0 \left({\bf p}-{{\bf q}\over 2} \right)
-\rho ^{(0)}_x \left({\bf p}+{{\bf q}\over 2} \right)
-\rho ^{(0)}_x \left({\bf p}-{{\bf q}\over 2} \right)
\over 2}
\ee
To obtain the susceptibility, we solve for
\be
\delta \rho _+ ({\bf q},\omega ) =
\int {d^3 p \over (2\pi )^3} \delta \rho _+ ({\bf p},{\bf q},\omega ),
\ee
which gives
\be
\label{ballistic chi}
\chi _{zz}^B (q, \omega >0)=
{1\over 4} \int {d^3 p \over (2\pi )^3}
{n\left( \xi \left({\bf p}-{{\bf q}\over 2}\right)-{\omega _L \over 2}\right)
-n\left( \xi \left({\bf p}+{{\bf q}\over 2}\right)+{\omega _L \over 2}\right)
\over
\omega - \omega _L - {{\bf p}\over m}\cdot {\bf q} +{i\over \tau _s}}.
\ee
The superscript $B$ denotes the case of ballistic motion.

For diffusive motion, the fermion momentum relaxes on a rapid time scale
$\tau << \tau _s$. In this case, the scattering contribution to the kinetic
equation is
\be
I_s=-{1\over \tau _s} \delta \rho ({\bf p},{\bf r},t)
-{1\over \tau}(\rho ({\bf p},{\bf r},t)-\rho _{eq}(p; \rho ({\bf r},t))),
\ee
where $\rho _{eq}(p; \rho ({\bf r},t))$ is the momentum equilibrium distribution
consistent with the local number-spin density $\rho ({\bf r},t)$,
\be
\int {d^3 p \over (2\pi )^3} \rho _{eq}(p; \rho ({\bf r},t))
=
\rho ({\bf r},t).
\ee
For $\delta \rho  ({\bf p},{\bf r},t)$ given by Eq.
\ref{deviations},
\be
I_s=-{1\over \tau _s} \delta \rho ({\bf p},{\bf r},t)
-{1\over \tau } \delta \rho  ({\bf p},{\bf r},t)
+{1\over \tau }
{n\left(\xi (p)-{\omega _L \over 2}\right)
- n\left(\xi (p)+{\omega _L \over 2}\right)
\over
\int {d^3 p \over (2\pi )^3}
\left(
n\left(\xi (p)-{\omega _L \over 2}\right)
- n\left(\xi (p)+{\omega _L \over 2}\right)\right)}
\delta \rho  ({\bf r},t).
\ee
To obtain the susceptibility at positive frequencies, we write
the kinetic equation for the circular component of the density matrix
\bea
-i\left( \omega -\omega _L-{{\bf p}\over m}\cdot {\bf q} +{i\over \tau _s}
+{i\over \tau} \right)
\delta \rho _+  ({\bf p},{\bf q},\omega)
& = & i \phi ({\bf q},\omega)
{
\rho ^{(0)}_0 \left({\bf p}+{{\bf q}\over 2} \right)
-\rho ^{(0)}_0 \left({\bf p}-{{\bf q}\over 2} \right)
-\rho ^{(0)}_x \left({\bf p}+{{\bf q}\over 2} \right)
-\rho ^{(0)}_x \left({\bf p}-{{\bf q}\over 2} \right)
\over 2}
\nonumber \\
& + &
{1\over \tau }
{n\left(\xi (p)-{\omega _L \over 2}\right)
- n\left(\xi (p)+{\omega _L \over 2}\right)
\over
\int {d^3 p \over (2\pi )^3}
\left(
n\left(\xi (p)-{\omega _L \over 2}\right)
- n\left(\xi (p)+{\omega _L \over 2}\right)\right)}
\delta \rho _+ ({\bf q},\omega).
\eea
To obtain the susceptibility, we solve for
\be
\delta \rho _+ ({\bf q},\omega ) =
\int {d^3 p \over (2\pi )^3} \delta \rho _+ ({\bf p},{\bf q},\omega ),
\ee
which to lowest order in $\tau $ gives
\be
\label{diffusive chi}
\chi _{zz}^D (q, \omega >0 ) = {1\over 4}
{
\int {d^3 p \over (2\pi )^3} \left(
n\left( \xi \left({\bf p}\right)-{\omega _L \over 2}\right)
-n\left( \xi \left({\bf p}\right)+{\omega _L \over 2}\right) \right)
\over
\omega - \omega _L +i (\tau _s^{-1} + D(\omega _L) q^2)}.
\ee
The superscript $D$ denotes the case of diffusive motion.
Here,
\be
D(\omega _L)= \tau
{\int {d^3 p \over (2\pi )^3} {p^2 \over 3m^2} \left(
n\left( \xi \left({\bf p}\right)-{\omega _L \over 2}\right)
-n\left( \xi \left({\bf p}\right)+{\omega _L \over 2}\right) \right)
\over
\int {d^3 p \over (2\pi )^3} \left(
n\left( \xi \left({\bf p}\right)-{\omega _L \over 2}\right)
-n\left( \xi \left({\bf p}\right)+{\omega _L \over 2}\right) \right)
}
.
\ee
In the limit $\omega _L \rightarrow 0$, $D(\omega _L)$ is equal to the usual
diffusion constant. For the classical distribution $n(\xi )$ in the
high-temperature limit, $D(\omega _L)$ is independent of $\omega _L$ for
any value of $\omega _L$. At $T=0$, the high-field limit $D(\omega _L/2>E_F)$ is
reduced compared to the low-field limit $D(\omega _L \rightarrow 0)$ by a
factor of $2^{2/3}\times 3/5=0.95$. Thus we consider $D$ independent
of the magnetic field, and drop the $\omega _L$ dependence in the following.

Substituting Eq. \ref{ballistic chi}
into Eq. \ref{FDthm} gives
\be
\label{SzzB}
S_{zz}^{B}(\omega ) \simeq {\pi \over 2}
\coth {\omega \over 2T} A\times L
\int {d^3 p \over (2\pi )^3}
\left[ {
n\left( \xi \left({\bf p}-{{\bf e}\over 2R}\right)-{\omega _L \over 2}\right)
-n\left( \xi \left({\bf p}+{{\bf e}\over 2R}\right)+{\omega _L \over 2}\right)
\over 2}\right]
{{1\over \pi \tau _s}
\over
\left(\omega - \omega _L - {{\bf p}\over m}\cdot {{\bf e}\over R}\right)^2
+{1\over \tau _s^2}},
\ee
for the case of the ballistic motion. Here, ${\bf e}$ is the unit vector in
an arbitrary direction.
Substituting Eq. \ref{diffusive chi}
into Eq. \ref{FDthm} gives
\be
\label{SzzD}
S_{zz}^{D}(\omega ) \simeq {\pi \over 2}
\coth {\omega \over 2T} A\times L
\int {d^3 p \over (2\pi )^3} \left[{
n\left( \xi \left({\bf p}\right)-{\omega _L \over 2}\right)
-n\left( \xi \left({\bf p}\right)+{\omega _L \over 2}\right)
\over 2}\right]
{{1\over \pi} (\tau _s^{-1} + D/R^2)
\over
(\omega - \omega _L)^2 + (\tau _s^{-1} + D/R^2)^2}
\ee
for the case of the diffusive motion.
Evaluating the integrals in Equations \ref{SzzB} and
\ref{SzzD}, see Appendix B,
gives results of the form of Eq. \ref{general formula}, with
specific forms for function
$f\left( {(\omega - \omega _L )/ g (t_{tr}^{-1}, \tau _s^{-1})} \right)$.
The forms of the function $f$ in the various cases are summarized in Table I.

In the diffusive case, the function $f$ is a Lorentzian for all temperatures, and
\be
g (t_{tr}^{-1},\tau _s^{-1})= t_{tr}^{-1} + \tau _s^{-1}.
\ee
The function $f$ in the ballistic case (\ref{ballistic chi}) is
more complicated, so we consider various limiting cases.
In the limit $\tau _s^{-1}>> v/R$, where $v$ is the Fermi velocity at
$T=0$ and the thermal velocity $\sqrt{2T/m}$ at $T>E_F$, we can drop
${\bf p}/m \cdot {\bf q}$ from the denominator of (\ref{ballistic chi}),
so the numerator becomes $q$ independent, and we obtain a
Lorentzian with the width $\tau _s^{-1}$, like in the diffusive regime
for $\tau _s^{-1}>>D/R^2$. In the limit of $\tau _s^{-1}$ dominating
the inverse travel time through the probed region, the spin noise cannot
distinguish between the ballistic motion and diffusive motion.
We can distinguish between these cases in the opposite limit
$\tau _s^{-1}\rightarrow 0$. Then the line
in the diffusive regime is still a Lorentzian, but now with the width
$D/R^2$. In the ballistic case, at $T=0$, the line has a trapezoidal
shape for $\omega _L<<E_F$ and a parabolic shape for $\omega _L >>E_F$.
In the classical limit $T>>E_F$, the line has a Gaussian shape.

{\Large
\begin{table}
\begin{tabular}{|l||l|l|}
\hline
&$R<v_s \tau $  & $R>v_s \tau $  \\
& (ballistic) & (diffusive)\\
\hline \hline
$T<E_F$
& Trapezoidal (for $\omega _L/2 < E_F$ )  &
Lorentzian \\
(degenerate) & Parabolic (for $\omega _L/2 > E_F$) &
\\
\hline
$T>E_F$ & Gaussian
& Lorentzian \\
(classical) & &
\\
\hline
\end{tabular}
\caption{Summary of the shapes of the noise spectral lines in the four regimes
in the limit $\tau _s^{-1}\rightarrow 0$.}
\label{Table}
\end{table}
}

\section{Conclusions}
\label{Conclusions}

Motivated by spin noise spectroscopy measurements,
we developed a theory of spin noise of itinerant fermions in different
regimes of temperature and disorder. We found a general result with a clear
physical interpretation, and showed how it follows from spin kinetic equations.
Our theory provides a framework for interpreting recent experiments on
atomic gases and conduction electrons in semiconductors and provides a baseline for
identifying the effects of interactions on spin noise spectroscopy.

\section{Acknowledgement}
This work was supported in part by the US Department of Energy, BES, and by the Los Alamos National Laboratory
LDRD program. 
Work in Cambridge was supported by EPSRC.
This work was also partially supported by the Ministry of Education
of the Czech Republic through Project No. MSM 4977751302.
We are grateful to S. A. Crooker and
D. E. Khmelnitskii for discussion and carefully
reading the manuscript.

\section*{Appendix A}

In this Appendix, we derive the spin noise power spectrum $S_{zz}(\omega)$
for a single spin $J$ in the magnetic field applied in the $x$ direction:
\bea
S_{zz}(t_2-t_1) & \equiv &
{1\over 2} \langle \{J_z(t_2), J_z(t_1) \} \rangle
\nonumber \\
& = &
{1\over 8}(e^{-i\omega _L (t_2-t_1)}+ e^{i\omega _L (t_2-t_1)})
\langle \{J_+, J_- \} \rangle
\eea
where
\be
J_{\pm }=J_z \mp iJ_y.
\ee
The thermal expectation value of the anticommutator is related to the thermal
expectation value of the commutator
\be
\langle \{J_+, J_- \} \rangle
= \coth {\omega _L \over 2T} \langle [J_+, J_- ] \rangle .
\ee
Using the commutation relation
\be
[J_+,J_-]=2J_x,
\ee
we arrive at the result
\be
\label{spinJnoise}
S_{zz}(\omega >0)= {\pi \over 2} \coth {\omega _L \over 2T}
\langle J_x \rangle
\delta (\omega - \omega _L).
\ee
We see that Eq. \ref{spinJnoise} has the form of Eq. \ref{general formula}
with
\be
{1 \over g (t_{tr}^{-1}, \tau _s^{-1})}
f\left( {\omega - \omega _L \over g (t_{tr}^{-1}, \tau _s^{-1})} \right)
= \delta (\omega - \omega _L).
\ee
The formula for magnetization
\be
\langle J_x \rangle =
{2J + 1\over 2}
\coth {(2J + 1) \omega _L \over 2T}
-
{1\over 2} \coth {\omega _L \over 2T}
\ee
implies that in the special case of $J=1/2$, the temperature and field dependence
cancels between $\coth \omega _L / 2T$ and $\langle J_x \rangle$ giving
\be
S_{zz}(\omega >0)= {\pi \over 4}
\delta (\omega - \omega _L).
\ee

\section*{Appendix B}

We present the details of calculations for the ballistic case with
$\tau _s^{-1}\rightarrow 0$ in three different limiting cases.

\subsection{$T=0$, $\omega _L<<E_F$}

In this case, we can linearize around the Fermi surface. From
(\ref{ballistic chi}), we now get
\bea
\chi _B''(q, \omega > 0)
& = &
-{\pi \over 4} N_0 \int d\xi \int\limits_{-1}^1 {dc\over 2}
\left( \theta \left(-\xi +{v_F q c + \omega _L \over 2}\right)
- \theta \left(-\xi -{v_F q c + \omega _L \over 2}\right) \right)
\delta(\omega - ( v_F q c + \omega _L ))
\nonumber \\
& \simeq & -{\pi \over 2}\times {N_0 \omega _L \over 2}
\times {\omega \over \omega _L}
{\theta (v_F q - |\omega - \omega _L|)\over 2v_F q}.
\eea
We thus obtain (\ref{general formula}) with the magnetization proportional to
the magnetic field via the Pauli susceptibility,
$M_x^{(0)}=N_0 \omega _L/2$, ballistic transport time
$t_{tr}=R/v_F$, and $f$ with
 a trapezoidal shape.

\subsection{$T=0$, $\omega _L/2>E_F$}

In this case, the fermions are fully polarized, so
\bea
\chi _B''(q, \omega > 0)
& = &
-{\pi \over 4}\int\limits _0^{\infty } {p^2 dp \over (2\pi )^2}
\int\limits_{-1}^1 dc \
\delta \left( \omega - \omega _L - {p \over m}qc \right)
\theta \left( \mu + {1\over 2}\left( \omega _L - {p \over m}qc \right)
-{p^2 + (q/2)^2 \over 2m}\right)
\nonumber \\
& \simeq &
-{1\over 16\pi }{m^2 \over q}
\int\limits _{{(\omega - \omega _L )^2 m \over 2q^2}} ^{\infty}
d\epsilon \  \theta \left( \mu + {\omega \over 2} - \epsilon \right),
\eea
where we neglected $q^2/2m$ compared to the Fermi energy $\mu + \omega _L/2$.
Using the formula for particle density
\be
n = {1\over 6 \pi ^2} p_F^3
\ee
with
\be
p_F=\sqrt{2m \left( \mu + {\omega _L \over 2} \right)},
\ee
we find
\be
\chi _B''(q, \omega > 0)
= -{\pi \over 2} \times {n\over 2} \times {3\over 4}{1\over qp_F/m}
\left[ 1- \left({\omega - \omega _L \over qp_F/m} \right)^2 \right].
\ee
The magnetization is now saturated at one half times the number of the
fermions in the probed region. The susceptibility is non-zero because
it is transverse to the external magnetic field ($zz$ response function with
the external magnetic field applied in the $x$ direction).
The transport time is still equal to $R/v_F$.
The line shape $f$ is now parabolic.

\subsection{$T>E_F$}

In this case, we approximate
\be
n(\xi ) = e^{-(\xi - \mu)/T}
\ee
with $\mu < 0$, so (\ref{general formula}) gives
\bea
\chi _B''(q, \omega > 0)
& = &
-{\pi \over 8}n
{\int\limits _0 ^{\infty} p^2 dp \int\limits _{-1}^1 dc
e^{-{p^2+(q/2)^2 \over 2mT}}
\sinh \left({pqc/m-\omega _L \over 2T}\right)
\delta (\omega - ( {pqc/m-\omega _L}))
\over
\int\limits _0 ^{\infty} p^2 dp
e^{-{p^2 \over 2mT}}\cosh {\omega _L \over 2T}}
\nonumber \\
& \simeq &
-{\pi \over 2} \times {n\over 2} \tanh {\omega _L \over 2T} \times
{e^{-{(\omega - \omega _L)^2 \over 2Tq^2/m}} \over \sqrt{2\pi Tq^2/m}}.
\eea
Thus, the equilibrium magnetization is $n/2 \tanh {\omega _L / 2T}$, the transport
time is $R/v$, where $v$ is the thermal velocity $\sqrt{2T/m}$,
and the line shape $f$ is Gaussian.


\begin{thebibliography}{99}
\bibitem{Sleator}T. Sleator, E.L. Hahn, C. Hilbert, and J. Clarke, Phys. Rev.
Lett. {\bf 55}, 1742 (1985).
\bibitem{Weissman}M.B. Weissman, Rev. Mod. Phys. {\bf 65}, 829 (1993).
\bibitem{Smith}N. Smith and P. Arnett, Appl. Phys. Lett. {\bf 78}, 1448 (2001).
\bibitem{dot}S.A. Crooker, J. Brandt, C. Sanfort, A. Greilich, D.R. Yakovlev,
D. Reuter, A.D. Wieck, M. Bayer, arXiv:0909.1592 (2009).
\bibitem{Aleksandrov}E.B. Aleksandrov and V.S. Zapassky, Zh. Eksp. Teor. Fiz.
{\bf 81}, 132 (Sov. Phys. JETP {\bf 54}(1), 64) (1981).
\bibitem{Mitsui}T. Mitsui, Phys. Rev. Lett. {\bf 84}, 5292 (2000).
\bibitem{noise nature} S.A. Crooker, D.G. Rickel, A.V. Balatsky, and D.L. Smith, Nature (London) {\bf 431}, 49 (2004).
\bibitem{Sorensen}J.L. Sorensen, J. Hald, and E.S. Polzik, Phys. Rev. Lett.
{\bf 80}, 3487 (1998).
\bibitem{Kuzmich}A. Kuzmich, L. Mandel, J. Janis, Y.E. Young, R. Ejnisman,
and N.P. Bigelow, Phys. Rev. A {\bf 60} 2346 (1999).
\bibitem{Oestreich} M. Oestreich, M. R{\" o}mer, R.J. Haug, and D. H{\" a}gele,
Phys. Rev. Lett. {\bf 95}, 216603 (2005).
\bibitem{nGaAsbulk}S.A. Crooker, L. Cheng, and D.L. Smith,
Phys. Rev. B {\bf 79}, 035208 (2009).
\bibitem{GaAsQW}G.M. M{\" u}ller, M. R{\" o}mer, D. Schuh, W. Wegscheider,
J. H{\" u}bner, and M. Oestreich,
Phys. Rev. Lett. {\bf 101}, 206601 (2008).
\bibitem{STM1}Y. Manassen, R.J. Hamers, J.E. Demuth and A.J. Castellano, Jr., Phys. Rev. Lett. {\bf 62}, 2531 (1989).
\bibitem{STM2}A. V. Balatsky, Y. Manassen and R. Salem, Phys. Rev. B {\bf 66}, 195416 (2002).
\bibitem{review}M. R{\" o}mer, J. H{\" u}bner, and M. Oestreich,
Rev. Sci. Instrum. {\bf 78}, 103903 (2007).
\bibitem{doping}M. R{\" o}mer, J. H{\" u}bner, and M. Oestreich,
Appl. Phys. Lett. {\bf 94}, 112105 (2009).
\end{thebibliography}
\end{document}